# A Photonic Integrated Circuit based Compressive Sensing Radio Frequency Receiver

David B. Borlaug,[1]* Steven Estrella,[2] Carl Boone,[1] George Sefler,[1] Thomas Justin Shaw,[1]
Angsuman Roy,[2] Leif Johansson,[2] George C. Valley[1]

[1]The Aerospace Corporation, 2310 E. El Segundo Blvd., El Segundo, CA 90245, USA
[2]Freedom Photonics, 41 Aero Camino, Santa Barbara, CA 93117, USA
*Corresponding author: david.b.borlaug@aero.org

**Abstract:** A photonic integrated circuit comprised of an 11 cm multimode speckle waveguide, a 1x32 splitter, and a linear grating coupler array is fabricated and utilized to receive 2 GHz of RF signal bandwidth from 2.5 to 4.5 GHz using a 35 MHz mode locked laser.

## Introduction

Compressive sensing (CS) enables the capture of multiple narrow-band signals that sparsely occupy broad bandwidth domains while sampling below the Nyquist rate. In this work, 16 radio-frequency (RF) channels, each with an effective sample rate of 35 MSps, are used to recover RF signals across 2 GHz of RF bandwidth from 2.5 to 4.5 GHz. Compared with direct Nyquist sampling which requires 9 GSps, the CS system required 9000 / (35 x 16) = 16 times less recorded samples and a maximum sampling rate 9000 / 35 = 257 times smaller. Previous works have demonstrated RF signal recovery [1-4] and optical reservoir computing [5] using the speckle of a multimode optical fiber. The speckle of a multimode waveguide in a photonic integrated circuit (PIC) was previously utilized in demonstrating an RF spectrometer [4, 6]. To the authors' knowledge, this work presented here demonstrates for the first time the use of waveguide speckle for the compressive sensing of RF signals in a compact PIC.

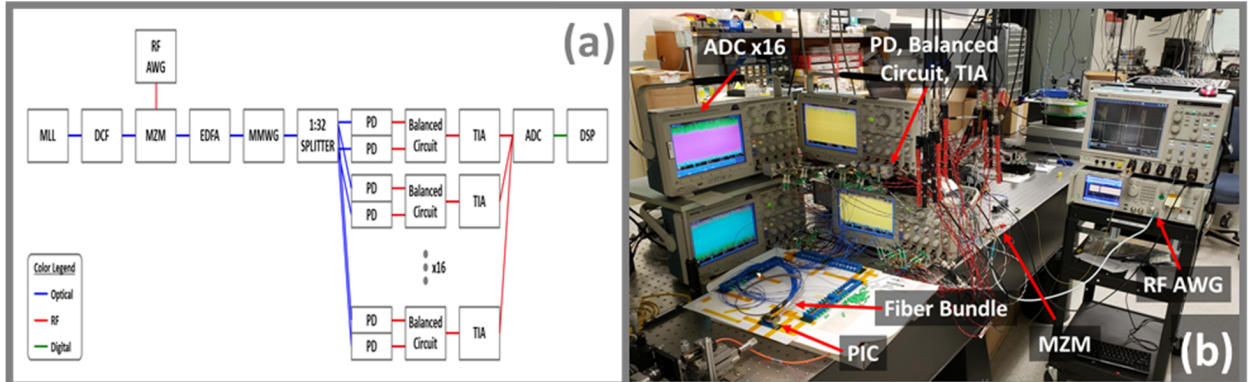

*Figure 1: (a) Experimental schematic. MLL: mode locked laser; DCF: dispersion compensating fiber; MZM: Mach Zehnder modulator; AWG: arbitrary waveform generator; EDFA: erbium doped fiber amplifier; MMWG: multimode waveguide; PD: photodiode; TIA: transimpedance amplifier; ADC: analog to digital converter; DSP: digital signal processor. (b) Photograph of experimental setup.*

**Experimental Setup.** The experimental setup is shown in Figure 1. The setup is the same as in [3] except that the multimode fiber, imaging apparatus, and 32-core fiber bundle in [3] have been replaced with the single PIC discussed in this work. In brief, a 35.71 MHz rep-rate MLL pulse is broadened to 4.5 ns in a DCF fiber, modulated with an RF AWG, amplified, and input to the PIC. The PIC hosts the multimode waveguide and the 1:32 splitter used for spatially sampling the multimode waveguide's speckle pattern. The photodiodes, balanced circuits, TIAs, ADCs, and DSP discussed in [3] remain unchanged.

**PIC Design, Fab, Package, Test.** Light is coupled to and from the PIC using a single linear fiber array with 48 populated fibers (c.f. Figure 1(b), Figure 2, Figure 3). The

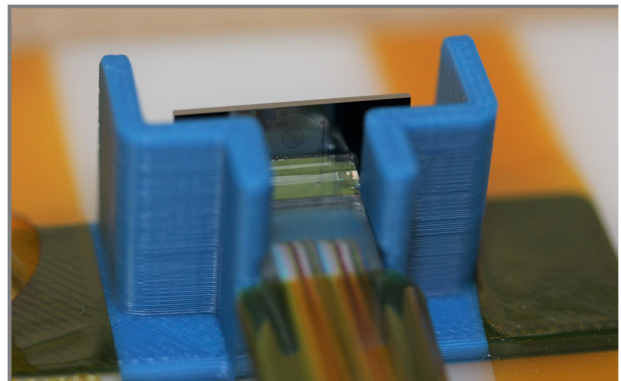

*Figure 2: The 6x6 mm PIC is attached to a 1x48 V-groove fiber array using UV adhesive. The fiber array is bedded in a 3D printed holder, with the PIC suspended by the adhesive. Note the multimode spiral waveguide.*

PIC hosts 44 grating couplers, 6 of which are devoted to fiber array alignment aids, 6 for optical input (only one used at any given time), and 32 for optical output. The alignment aids consist of 2 beacon waveguides that emit light out of plane for imaging by an overhead camera during first-light rough alignment, and 2 loop-back couplers for automated precision alignment. The count of the 32 optical outputs are designed to match the testbed described in [3], making the PIC a drag and drop replacement for the multimode fiber, imaging system, and 32-core fiber bundle. The PIC uses passive silicon on insulator technology with 220 nm thick silicon, a 75 nm shallow etch, and a 220 nm through etch. The grating couplers utilize an adiabatic taper and focusing grating with 628 nm pitch, 50 percent duty cycle, and uniform period.

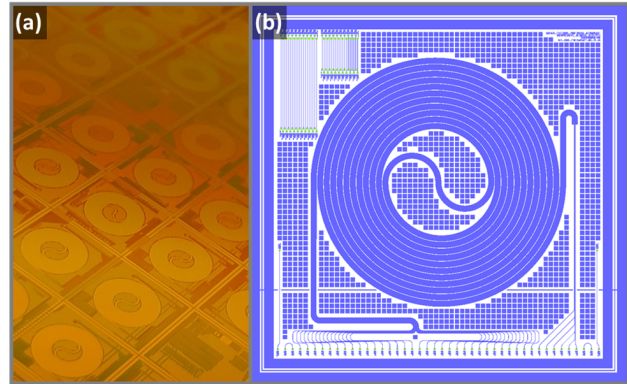

*Figure 3: (a) A micrograph of wafer-level photolithographic production of compressive sensing receivers, and (b) the design artwork for the 6x6 mm compressive sensing receiver.*

The 6 single-mode input waveguides on the right side of the spiral (c.f. Figure 3) excite the 11 cm long 96 μm wide multimode spiral speckle waveguide. The speckle mixer output is spatially sampled on the PIC into 32 single-mode waveguides, and each waveguide is routed to a fiber grating coupler. A series of "trombone" matched length delay lines on the output path ensures each waveguide experiences the same delay during routing. The grating couplers were measured to have 3.8 dB insertion loss per coupler. The 1:32 split accounts for a 15 dB reduction in output channel power. The multimode waveguide and on-chip routing account for approximately 13 dB of loss, or 1.2 dB per cm. However, a shorter multimode waveguide may be sufficient to produce the required speckle with less loss. The 6x6 mm PIC artwork and fabrication process was designed at The Aerospace Corporation, fabricated by Freedom Photonics, and packaged by PLC Connections. The system testbed and data analysis are discussed in detail in [3].

**Results.** The procedure for performing the compressive sensing calibration, reconstructing the training dictionary, and minimizing the penalized $l_1$ norm to recover RF amplitude, phase and frequency is detailed in [3]. Due to optical loss in the PIC, averaging over 512 laser pulses was performed. Figure 4(a) shows successful reconstruction of the in-phase and quadrature components of the training dictionary at 57-frequencies between 2.55 and 4.51 GHz. Figure 4(d) shows the successful reconstruction of the 100 μs training dictionary. In the following 200 μs of Figure 4(d), two tones are simultaneously excited and successfully reconstructed, one static at 3.5 GHz, and one swept from 2.55 GHz to 4.51 GHz. Future PIC designs may be improved by directly integrating photodiodes on chip to avoid grating coupler and waveguide routing losses. On-chip photodetection would also alleviate channel-count limitations imposed by the V-groove fiber array's physical size and enable higher spatial sampling of the waveguide speckle pattern. A chirped waveguide Bragg grating may also replace the DCF for higher pulse repetition rates. High speed modulators are available in both silicon and InP platforms. The InP platform opens the possibility for developing a PIC-based high rep-rate MLL followed by an integrated pulse picker. InP enables semiconductor optical amplifiers prior to modulation and immediately preceding photodetection. These techniques may enable improved frequency coverage and improve the system signal-to-noise ratio.

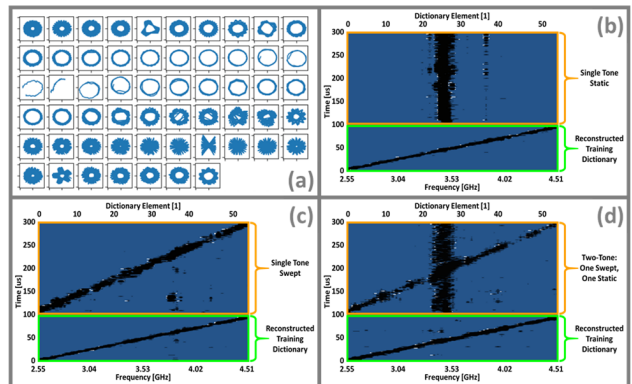

*Figure 4: (a) Successful recovery of dictionary signal phase (I & Q) using compressive sensing (2.55 GHz top-left, to 4.51 GHz bottom-right). (b) Reconstructed dictionary and static single-tone test. (c) Swept single-tone test. (d) dual-tone test with one static tone and one swept tone.*

**Acknowledgment.** This work was supported by NAVAIR Contract Number N68335-18-C-0089.